\documentclass[prb,twocolumn,showpacs,aps]{revtex4}
\usepackage{amssymb}
\usepackage{amsmath}
\usepackage[dvips]{graphicx}
\usepackage{color}
\begin{document}

\title{Thermally driven pure spin and valley current via anomalous Nernst effect in monolayer group-VI dichalcogenides}
\author{Xiao-Qin Yu$^{1,2}$}

\author{Zhen-Gang Zhu$^{1,3}$}
\email{zgzhu@ucas.ac.cn}

\author{Gang Su$^{3}$}
\email{gsu@ucas.ac.cn}

\author{A. -P. Jauho$^{4}$}
\email{Antti-Pekka.Jauho@nanotech.dtu.dk}

\affiliation{$^{1}$School of Electronic, Electrical and Communication Engineering, University of Chinese Academy of
Sciences, Beijing 100049, China. \\
$^{2}$Sino-Danish center for Education and Research, University of Chinese Academy of Sciences, Beijing 10049, China.\\
$^{3}$ Theoretical Condensed Matter Physics and Computational Materials Physics Laboratory, College of Physical Sciences, University of Chinese Academy of Sciences, Beijing 100049, China.\\
$^{4}$Center for Nanostructured Graphene (CNG), DTU Nanotech, Department of Micro- and Nanotechnology, Technical University of Denmark, DK-2800 Kgs. Lyngby, Denmark.}

\begin{abstract}
Spin and valley dependent anomalous Nernst effect are analyzed for monolayer MoS$_2$ and other group-VI dichalcogenides. We find that pure spin and valley currents can be generated perpendicular to the applied thermal gradient in the plane of these two-dimensional materials. This effect provides a versatile platform for applications of spin caloritronics. A spin current purity factor is introduced to quantify this effect. When time reversal symmetry is violated, e.g.  two-dimensional materials on an insulating magnetic substrate, a dip-peak feature appears for the total Nernst coefficient. For the dip state it is found that carriers with only one spin and from one valley are driven by the temperature gradient.

\end{abstract}
\maketitle
%\setlength{\parindent}{8pt}

%\section{Introduction}
The generation of a spin current is a vital issue in spintronics \cite{Wolf S A,Igor,David,Albert}
in which information is carried and stored by spin, rather than charge. A traditional approach to generate a spin current is to drive a spin polarized charge current through a spin filter \cite{Wolf S A,Igor,Amnon,Z.-G.Zhu}, where a ferromagnet is used as the source of spin. However, the net charge current may lead to dissipation in the device, and the introduction of a ferromagnet - semiconductor interface complicates the applications. To eliminate the dissipation and complexity, pure spin current generation in semiconductors without a net charge current has been recently explored. 
Spin Hall effect \cite{Hirsch,S.Murakami,N.A.Sinitsyn,David,Y.K.Kato} 
has been proposed in semiconductors with strong spin-orbit coupling (SOC),
which describes the generation of dissipationless spin current in the perpendicular direction (y-direction) to an applied electric field (x-direction) without breaking time-reversal symmetry. It can be attributed to a nonvanishing Berry curvature of energy bands in the presence of such an external electric field.

Recently, a parallel research field to spintronics was introduced: spin caloritronics \cite{G.E,A.D.Avery,S.Y.Huang,K.Uchida,J.Xiao,H.Adachi}, which is an extension and combination of spintronics and the conventional thermoelectrics, aiming at increasing the efficiency and versatility of spin-involved thermoelectric devices \cite{S.T.B}.
So far, the spin Seebeck effect \cite{K.Uchida1,K.Uchida2,K.Uchida3,C.M.Jaworski} was suggested to be a way for spin current generation as a consequence of a temperature gradient. Usually a net charge current is still present in loop circuit use.

Monolayer  MoS$_{2}$ and other transition metal dichalcogenides (TMDC) represent a new class of two-dimensional (2D) materials, intrinsically behaving as semiconductors. Analogous to gapped graphene \cite{K.S2,K.S3}, their Berry curvature in each Dirac cone is nonvanishing and leads to a series of anomalous transport phenomena, such as the anomalous Hall effect \cite{N.Nagaosa}, valley Hall effect, anomalous Nernst effect (ANE) and valley ANE. However, unlike graphene, inversion symmetry is broken in TMDCs and they exhibit strong SOCs, which lead to coupling of the spin and valley degrees of freedom. Xiao \textit{et al}. have studied the anomalous Hall effect and the related spin and valley Hall effect \cite{D.Xiao}. Mak \textit{et al.} have observed a valley Hall effect in MoS$_{2}$ transistors \cite{K.F.Mak}. 
The strong SOC  offers a great chance for spintronics and spin caloritronics. Recently, Fang \textit{et al.} \cite{fang} found an extrinsic mechanism via spin Seebeck effect to produce pure spin current through temperature gradient in ferromagnetic/nonmagnetic hybrid metallic system.
Here, we propose a new intrinsic mechanism to generate
{\it a pure spin current} via spin Nernst effect (SNE) driven by a temperature gradient 
in monolayer of MoS$_2$ and other TMDCs. Since the transistor based on MoS$_{2}$ has been fabricated and has high equality of performance, the proposed effect is highly applicable in reality. 
In addition, the pure spin current persists for a wide range of gate voltage (for example, 0.42 eV for WS$_2$). We believe that this effect is very useful in spin caloritronics.

The effective Hamiltonian of MoS$_{2}$ around Dirac cones is \cite{D.Xiao}
\begin{equation}
\hat{H}=at({\tau}k_x \hat{\sigma}_x+k_y \hat{\sigma}_y)+\frac{\Delta}{2}\hat{\sigma}_z-\lambda{\tau}\frac{\hat{\sigma}_z-1}{2}\hat{s}_z,
\label{hamiltonian}
\end{equation}
where $\tau={\pm}1$ is the valley index, $2\lambda$ refers to the spin splitting at the top of valence band caused by the SOC, $\hat{\boldsymbol{\sigma}}$ denotes the Pauli matrices for the two basis functions of the energy bands, $a$ is the lattice constant, $t$ is the hopping integral, $\Delta$ is the energy gap, and $\hat{\boldsymbol{s}}$ represent Pauli matrices for spin. The energy eigenvalues are
\begin{equation}
E_{n\tau s_z}=s_z\frac{\lambda{\tau}}{2}+n \sqrt{(kat)^2+\left(\frac{\Delta-s_z\lambda{\tau}}{2}\right)^2},
\end{equation}
where $s_z(=\pm1$) indicates the spin index, and $n(={\pm}1$) is the band index.
The Berry curvature \cite{N.Nagaosa,D.Xiao2,D.Xiao3,Z.-G} is determined by $\Omega_{n\tau s_z}(\mathbf{k})=\hat{z}\cdot\nabla_{\mathbf{k}}\times\langle{\mu_{n{\tau} s_z}}| i\nabla_{\mathbf{k}}|{\mu_{{n\tau} s_z}}\rangle$ for 2D materials, where $\nabla_{\mathbf{k}}$ means directional derivatives with respect to the momentum $\mathbf{k}$, and $\mu_{n{\tau} s_z}$ is the periodic part of the Bloch function.
For massive Dirac fermions described by the effective Hamiltonian in Eq. (\ref{hamiltonian}), the Berry curvature is
\begin{equation}
\Omega_{n {\tau}s_z}(\mathbf{k})=-{\tau}n \frac{2a^2t^2\Delta'}{[(\Delta')^2+4(kat)^2]^\frac{3}{2}},
\end{equation}
where $\Delta'=\Delta-s_z\lambda\tau$.

When a temperature gradient is applied, an electric field develops in the opposite direction due to the Seebeck effect. Besides the parallel effect, in the presence of the temperature gradient, the holes (or electrons) experience a Lorentz-like force and thus move in the direction perpendicular to the diffusion current, which is the ANE induced by the intrinsic non-vanishing Berry curvature \cite{J.M.Ziman}. This is also the intrinsic mechanism contributing to the anomalous Hall resistivity \cite{N.Nagaosa,D.Xiao3}. Thus, the velocity multiplied by the entropy density gives rise to the anomalous Nernst coefficient (ANC) \cite{Zhang C1,zhang C2,Z.-G Zhu1,Q.H.Wang}(details can be found in Ref. \onlinecite{SM}) in each Dirac cone with a specified spin
\begin{equation}
\alpha^{\text{c(v)}}_{{\tau}s_z}=4\pi\alpha_0\int\frac{d^2\mathbf{k}}{(2\pi)^2}\Omega_{n{\tau} s_z}(\mathbf{k})S_{n{\tau} s_z}(\mathbf{k}),
\label{nc}
\end{equation}
where c(v) represents the conduction(valence) band, corresponding to $n=1(-1)$, respectively. $\alpha_0=\frac{ek_B}{2h}$, $e$ is the electron charge, $h$ is the Planck constant, % $\tau=\pm1$ for K and -K, $n=+1 (-1)$ for the conduction (valley) bang,
$S_{n{\tau} s_z}(\mathbf{k})=-f_{n {\tau}s_z\mathbf{k}}\ln f_{n {\tau}s_z\mathbf{k}}-\left(1-f_{n{\tau} s_z\mathbf{k}}\right)\ln\left(1-f_{n{\tau}s_z\mathbf{k}}\right)$ ($k_B$ has been taken into $\alpha_0$)  is the entropy density for valley $\tau$, spin $s_z$, and $n$ band, $k_B$ is the Boltzmann constant, and $f_{n{\tau} s_z\mathbf{k}}$ is the Fermi distribution function. The entropy density develops a peak at $E=E_f$, and is essentially zero when the energy is beyond the range of $\left[E_f-5k_BT, E_f+5k_BT\right]$. The integration is performed over the neighborhood of one $K (-K)$ point in the momentum space. $\Omega_{n{\tau}s_z}(\mathbf{k})$ and $S_{n{\tau}s_z}(\mathbf{k})$ can be expressed as functions of the modulus of the wave vector {\bf k}, and we can use the Debye model, namely, the integration is taken in a circular region centered at $K (-K)$ point and the area is equal to the half of the first Brillouin zone for $K (-K)$ cone. Thus, Eq. (\ref{nc}) can be written as
\begin{equation}
\alpha^{\text{c(v)}}_{{\tau}s_z}=2\alpha_0\int^{k_{c}}_{0}\Omega_{n{\tau} s_z}(k)S_{n{\tau} s_z}(k)kdk,
\label{nc1}
\end{equation}
where $k_{c}=\frac{2\sqrt{\pi}}{3^{3/4}a}$. The ANC for the $\tau$ valley is then
\begin{equation}
\alpha^{\text{valley}}_{n{\tau}}=2\alpha_0\int^{k_{c}}_0[\Omega_{n{\tau},\uparrow}(k)S_{n{\tau},\uparrow}(k)
+\Omega_{n{\tau},\downarrow}(k)S_{n{\tau},\downarrow}(k)]kdk.
\label{vnc}
\end{equation}
The spin Nernst coefficient (SNC)  reads
\begin{equation}
\alpha^{\text{spin}}_{n{\tau}}=2\alpha_{0}^{s}\int^{k_{c}}_0[\Omega_{n{\tau},\uparrow}(k)S_{n{\tau},\uparrow}(k)
-\Omega_{n{\tau},\downarrow}(k)S_{n{\tau},\downarrow}(k)]kdk,
\label{spinnc}
\end{equation}
where $\alpha_{0}^{s}=\frac{\alpha_0\hbar}{2e}=\frac{k_B}{8\pi}$.

 \begin{figure}[tb]
 \centering
\includegraphics[width=0.94\linewidth,clip]{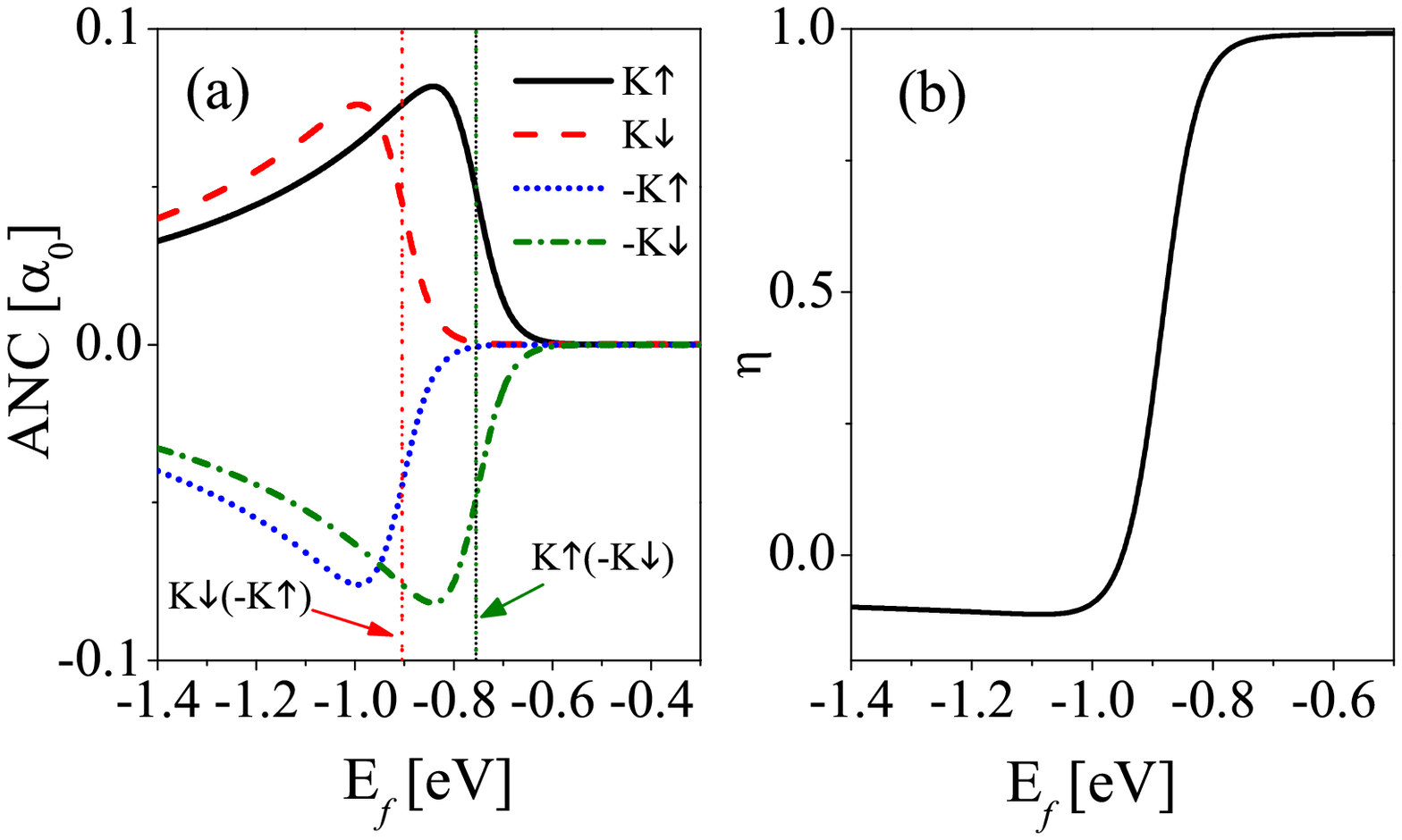}
\includegraphics[width=0.94\linewidth,clip]{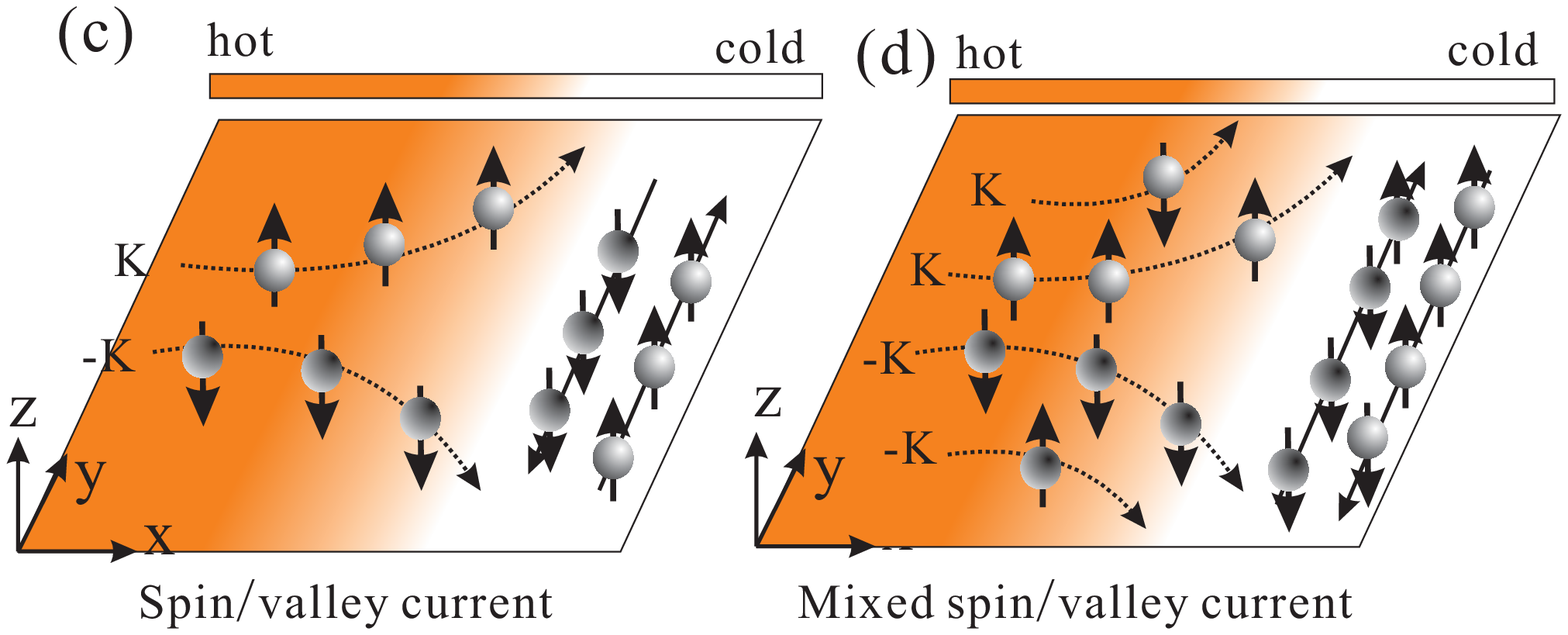}
\caption{(a) Illustration of the anomalous Nernst coefficient as a function of the Fermi energy for different spin states in each Dirac cone. The vertical dotted lines indicate the positions of the maxima  of the
valence bands with corresponding spin states and Dirac cones. (b) Spin current purity corresponding to the curves shown in (a). (c) A pure spin current and valley current can be generated when lowering the Fermi level into the bands of $K\uparrow$ ($-K\downarrow$). (d) Schematic illustration for the corresponding mixed spin (valley) current when further lowering the Fermi level into $K\downarrow$ ($-K\uparrow$). The 2D material is chosen as MoS$_2$ and the temperature is T=300 K.}
\label{1}
\end{figure}

The magnitude of the ANE is determined by the ANC. In the following, we first investigate the properties of the ANC for the valley $K$ for a free standing layer, followed by a discussion where a magnetic substrate is considered. %For another valley, it's the corresponding time reversal.
Based on the material parameters (Table \ref{parameters})\cite{D.Xiao}, %which are derived by fitting the results from the first-principles calculation,
the ANC was numerically calculated for the spin Nernst effect and Valley Nernst effect.  %as a function of Fermi energy, respectively, for different temperatures and materials.
Fig.~\ref{1}(a) displays the ANC for each cone and each spin state.
The ANCs for the two spin states in a given valley have the same sign but they are shifted in opposite directions in the energy axis due to the  SOC in valence band ($E_f<0$). Since the state of $-K\downarrow(\uparrow)$ is degenerate to $K\uparrow(\downarrow)$, the corresponding ANCs have the same magnitude but opposite signs. This gives rise to a striking effect that a pure spin current and valley current can be generated when the Fermi level is lying in the valence band. According to the spin splitting determined by the energy gap and the SOC of the material, this nearly 100\% spin current can be generated in a sizable range of energies (Table \ref{parameters}). For example, for MoS$_{2}$, this energy range is around 0.11 eV. This region is larger for other TMDCs, e.g. WS$_{2}$. The pure spin current and valley current generation in this case are schematically shown in Fig. \ref{1}(c). With a further lowering of the Fermi level, the purity of the spin and valley current is reduced. To characterize the extent of such mixing, we define a spin current purity factor (SCPF)
\begin{equation}
\eta=\frac{(\alpha^{\text{v}}_{K\uparrow}+\alpha^{\text{v}}_{-K\uparrow})-(\alpha^{\text{v}}_{K\downarrow}
+\alpha^{\text{v}}_{-K\downarrow})}{\mid\alpha^{\text{v}}_{K\uparrow}\mid+
\mid\alpha^{\text{v}}_{-K\uparrow}\mid+\mid\alpha^{\text{v}}_{K\downarrow}\mid+\mid\alpha^{\text{v}}_{-K\downarrow}\mid},
\label{eta}
\end{equation}
where $\alpha^{\text{v}}_{K\uparrow(\downarrow)}$ is determined by Eq. (\ref{nc}) for K cone in valence band with up (down) spin.
When $\eta=\pm1$, a pure spin current is generated in the $y$-direction driven by a temperature gradient in the $x$-direction. Otherwise, there will be mixing from different valleys and spin states. The variation of $\eta$ for MoS$_2$ is shown in Fig. \ref{1}(b) and a schematic graph of the mixed spin and valley current is given in Fig. \ref{1}(d).
For MoS$_2$ the $\eta$ factor becomes ill-defined for Fermi level approximately above $-0.5$ eV ($E_f > -0.5$ eV), because the ANCs become negligibly small (see Fig. 1(a)), and in the subsequent discussion we do not consider these energy regimes.

%When the energy is above -0.6 eV, the $\eta$ factor is not really defined for MoS$_{2}$ since the ANC for each cone and each %spin is negligibly small ($\sim$$10^{-4}\alpha_{0}$). And the definition of $\eta$ would be meaningless.

\begin{table}[tbph]
%\begin{table}[tbh]
\centering
\caption{Parameters for TMDCs. The energy unit is eV for $\Delta$ and $\lambda$. The last two columns indicate the range of gate voltages $\Delta E_{\text{gate}}=E_{f1}-E_{f2}$, where $E_{f1}$ and $E_{f2}$ define the upper and lower limit in which spin current purity $\eta>90\%$ and $\eta>98\%$ can be generated, respectively.
%{\bf Besides the upper limit $E_{f1}$ is taken where $\alpha_{K\uparrow(\downarrow)}$ is nearly zero($\sim$$10^{-4}\alpha_{0}$} %)
}%
\begin{tabular*}{8 cm}{@{\extracolsep{\fill}}lcccc}
%\begin{tabular*}{17cm}{@{\extracolsep{\fill}}lcccccc}
 \hline\hline
 &$\Delta$ &$2\lambda$ &$\Delta E_{\text{gate}}$ ($\eta>90\%$) &$\Delta E_{\text{gate}}$ ($\eta>98\%$)\\
 MoS$_2$&1.66 &0.15 &0.195 eV &0.119 eV\\
 MoSe$_2$&1.47 &0.18 &0.232 eV &0.177 eV\\
 WS$_2$ &1.79 &0.43 &0.472 eV &0.421 eV\\
 WSe$_2$ &1.60 &0.46 &0.502 eV &0.454 eV\\
 \hline\hline
 \end{tabular*}
 \label{parameters}
 \end{table}

\begin{figure}[tb]
\centering
\includegraphics[width=1.0\linewidth,clip]{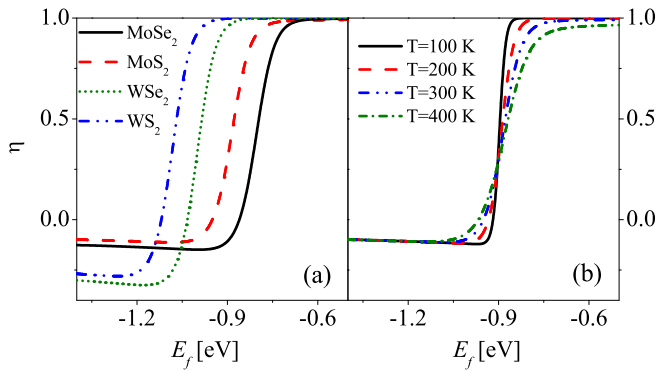}
\caption{(a) The spin current purity factor $\eta$  for the materials of Table 1 at $T=300$ K. (b) The temperature dependence of $\eta$ for MoS$_2$.}
\label{2}
\end{figure}

Fig. \ref{2} shows the SCPF for different materials and temperatures. An almost 100 \% spin current can be generated in a wider range of gate voltage for materials with larger spin splitting.
For example, in the energy scale from -0.6 eV to -1.0 eV for WS$_{2}$ (about 0.4 eV, which is quite large), the SCPF is almost 100\%.
%
%
%{\bf When the energy is above -0.6 eV, the SCPF is persistent to be nearly 100\%. However, $\alpha_{K\uparrow(\downarrow)}$ in %this case is tending to zero because the Fermi level is now in the energy gap.}
%{\bf I don't see this from Fig. 2!  How does eta behave at higher energies (above -0.6 eV)?  To see the entire range, one should also see that eta goes down, but this is not shown.}
At lower temperatures the SCPF is enhanced [Fig. \ref{2}(b)] as the thermal activation is suppressed. Another feature is that the SCPF changes its sign when the Fermi level is shifted sufficiently deep in the valence band, which means the generated spin current is reversed. In the deep valence band case, the W-based materials have larger $|\eta|$ (but negative, and can be as large as -0.3)  because of their stronger SOC.

%The appearing peaks are not exactly located at the top of the valence bands  but slightly higher than the top. The energy difference for the peaks is \textbf{proximately} equal to the spin splitting of the bands due to the strong SOC.
%
%\textbf{ with the peak appearing respectively slightly shift towards larger $|E_f|$ from the top of band for corresponding state}.
%
%And the energy difference for the peaks is equal to the spin-splitting due to the SOC.
%
%
%For the existence and the shift of NC peaks, it can be explained as follows. When $E_f$ is lowered from zero but still higher than the top of valence band, the NC increases owing to the the area of the entropy density in the momentum space increases. Continue lowering the $E_{f}$ into the valence band, the Berry curvature will be decreased, which weakens the effect of the larger entropy density.
%When these two effect exactly offset, the peak appears. A similar explanation can be applied to that of conduction band.

\begin{figure}[tb]
\centering
\includegraphics[width=1.0\linewidth,clip]{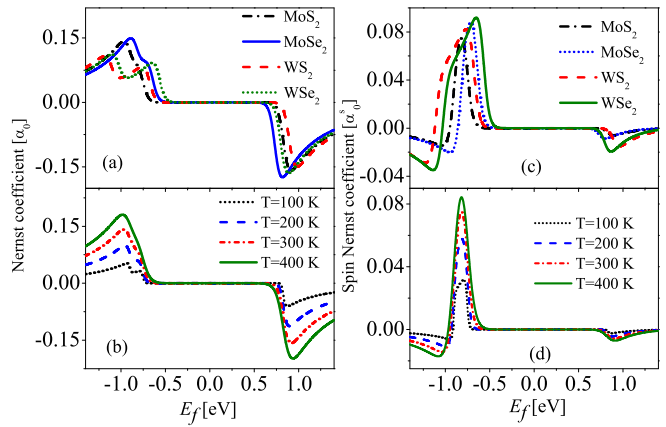}
\caption{The ANCs [(a), (b)] and SNC [(c), (d)] for K cone are calculated as a function of the Fermi energy for different materials and temperatures, respectively. The temperature in (a) and (c) is taken at 300 K. MoS$_2$ is fixed in (b) and (d).}
\label{3}
\end{figure}

The valley ANC is a sum of the two spin components and the SNC is a difference between them. Therefore, a two-peak feature of the valley ANC [Fig. \ref{3}(a)] and a dip-peak feature of the SNC of the valence band [Fig. \ref{3}(c)] can be observed due to the energy shifts of the two spin states.
The SNC undergoes a sign-change when lowering the Fermi level sufficiently. %Furthermore the spin NC and valley NC have the same magnitude in valence band before the spin-down states contribute to the NC ($E_{f}>-0.81$ eV).
The value of the valley ANC (SNC) for MoS$_2$ can reach 0.14$\alpha_{0}$ (0.07$\alpha_{0})$ at room temperature, which is comparable to the valley ANC in graphene \cite{Z.-G Zhu1}. The different behavior between MoS$_2$ and graphene is that the Nernst coefficients are no longer spin degenerate, leading to a large SNC for MoS$_2$. Nonzero SNC reflects that there is a spin imbalance at the opposite edges of the sample (open circuit case), which could be a source of spin injection and spin current generation (loop circuit case) in future applications of spin caloritronics. %Moreover it has established that the injected spin current into Pt can be converted to a usual voltage drop with designed orientations.
%A large SNC implies that the thermally driven spin current may be useful in future applications of spin caloritronics.

The valley ANC are strongly affected by the metal elements in the TMDCs, as shown in Fig.~\ref{3}(a). When the Fermi level lies in the valence band, the two-peak feature becomes more distinct with increasing of the SOC, as one moves through MoS$_{2}$, MoSe$_{2}$, WS$_{2}$ to WSe$_{2}$. The two peaks in the valence band can be explained by the spin splitting and the strong SOC. %{\bf due to the strong SOC ---delete}.
For instance, the energy difference between the two peaks of WSe$_{2}$ (WS$_{2}$) [Fig.~\ref{3}(a)] is 0.454 eV ( 0.43 eV), compatible to the spin splitting of the valence band 0.46 eV (0.45 eV). Thus, measuring the separation of the two peaks provides a method to estimate the effect of the spin splitting and the SOC.
The two-peak feature of the valley ANC becomes more distinct at low temperature since the peak of the entropy density around the Fermi level is sharper [Fig.~\ref{3}(b)]. Raising the temperature enhances the magnitude of the peak of the SNC but does not affect the positions (Fig.~\ref{3}(c)). %The peak positions of the SNC are not affected by temperature but are different for materials (Fig.~\ref{3}(c) and Fig.~\ref{3}(d)).
\begin{figure}[tb]
\centering
\includegraphics[width=0.75\linewidth,clip]{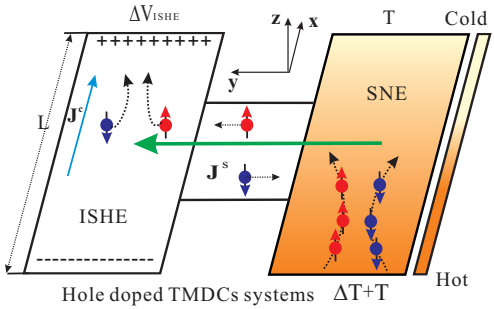}
\caption{A proposed H-shape detector for detecting the spin Nernst effect in TMDCs. The spin current, generated in the right leg by a temperature gradient, is injected into the left leg through a horizontal bridge, which can be converted into a detectable charge voltage drop $\Delta V_{\text{ISHE}}$ by the inverse spin Hall effect.}
\label{4}
\end{figure}

To make an estimation, $\alpha_{0}\approx1.665 \text{ nAK}^{-1}$. %corresponds to 50 $\mu \text{VK}^{-1}$ of a quantum resistivity $h/e^{2}$.
Thus, $\alpha^\text{spin}_{n}\approx0.08\times2\alpha_{0}\left(\frac{\hbar}{2e}\right)$ (Fig. 3(d), $2$ for valley degeneracy) for $\text{MoS}_{2}$ at room temperature corresponds to 0.54 nA charge current for $\Delta T=2$K, which should be detectable.
Specifically, to measure the SNE, we propose an H-shape detector \cite{brune} that is made of TMDCs (see Fig. \ref{4}). A temperature drop $\Delta T$ in $x$ direction is introduced in the right leg to generate spin current via the SNE: $\boldsymbol{J}^{s}=\alpha^{\text{spin}}_{n}(-\partial_{x}T) \mathbf{e}_{y}$. The spin current is injected into the left leg through a bridge that is supposed to be short (in ballistic regime) and narrow. On account of strong SOC in TMDCs, the electric field \cite{ando} $\boldsymbol{E}_{\text{ISHE}}=\rho\theta_{\text{SHE}}(\frac{2e}{\hbar}) \boldsymbol{J}^{s}\times\hat{\boldsymbol{s}}$ will be generated perpendicular to the spin direction $\hat{\boldsymbol{s}}$ (z direction here) via the inverse spin Hall effect (ISHE), resulting in a charge voltage drop (in open circuit case) $\Delta V_{\text{ISHE}}=-\frac{\sigma_{\text{SH}}}{\sigma^{2}}(\frac{2e}{\hbar})\alpha^{\text{spin}}_{n}\Delta T$, where $\theta_{\text{SHE}}=\sigma_{\text{SH}}/\sigma$ is spin Hall angle,   $\sigma_{\text{SH}}$ is spin Hall conductivity, and $\sigma$ is conductivity estimated by $n_{c} \mu e$. Mobility $\mu$ of MoS$_{2}$ is quite different in experiments. The early reported mobilities \cite{novoselovpnas2005,Q.H.Wang} range from $0.5$ to $3$ cm$^{2}$V$^{-1}$s$^{-1}$  for a MoS$_{2}$ transistor. It has been raised to $200$ cm$^{2}$V$^{-1}$s$^{-1}$ in a MoS$_{2}$ transistor with halfnium oxide gate recently \cite{Radisa}. Although MoS$_{2}$ p-type transistors have been fabricated successfully \cite{S.Chuang}, there are still lack of experimental data for $\sigma$. We exploit $\mu=400$ cm$^{2}v^{-1}s^{-1}$ and carrier density $n_{c}=10^{11}$ cm$^{-2}$ from Ref. \onlinecite{Kaasbjerg}. The $\sigma_{\text{SH}}=-1.16\pi\times10^{-2}\frac{e^{2}}{h}$ \cite{WX}. Therefore, $\sigma\approx16.544\times10^{-2}\frac{e^{2}}{h}$. Thus $|\Delta V_{\text{ISHE}}|\simeq 18.32\mu$V at room temperature for MoS$_{2}$ with $\Delta T$=2K, which should be measurable \cite{K.Uchida1}. Lower $\sigma$ (low $\mu$ and $n_{c}$) of the left leg leads to a larger voltage drop. The voltage drop is even larger for other TMDCs with stronger SOC.
\begin{figure}[t]
\centering
\includegraphics[width=0.9\linewidth]{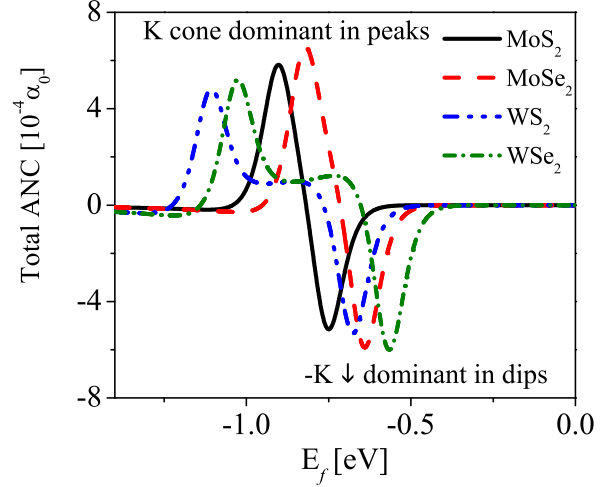}
\caption{The total ANC as function of the Fermi energy for different materials at 300K with the magnetization of magnet $M_z=6T$.}
\label{5}
\end{figure}

%\section{The ANC when applying insulating magnetic substrate}
Moreover, because of the time-reversal symmetry, no net ANE is produced. It is instructive to consider a situation where MoS$_{2}$ layer is placed on a magnetic insulating substrate with a perpendicular moment to resolve more intrinsic information of the spin and valley Nernst effect.
An insulating ferromagnetic yttrium iron garnet (YIG=Y$_3$Fe$_5$O$_{12}$) film could serve as such a substrate, in which %. Single crystalline (111) YIG layers are magnetically soft and isotropic in the film plane.  At room temperature, the YIG layer has
an inplane \cite{Y.M.Lu} %{ \color{blue}{The perpendicular magnetized YIG films had been realized experimentally\cite{W.X.Xiu}}}.
or a perpendicular magnetization \cite{W.X.Xiu} can be realized experimentally.
A large magnetic moment may be induced by a weak magnetic field so that the Landau level structure may be ignored. The Zeeman term, i.e. $-\frac{1}{2}g\mu_{B}M_{z}s_{z}$, should be added to Eq. (\ref{hamiltonian}). It has no impact on the eigenfunctions and the Berry curvature. Nonetheless it gives rise to  corrections to the eigenvalues, resulting in an asymmetry of the valence bands in different valleys. %Other ferromagnetic insulators may bring the same consequence to that of YIG.

In this circumstance, the time-reversal symmetry is broken so that the total ANC is nonvanishing, shown in Fig. \ref{5}. The typical feature is the dip-peak profile of the total ANC. At the $K$ point the down (up) spin band is shifted upward (downward);  while in the $-K$ valley, the up (down) spin band is shifted upward (downward). The dips are dominated by $-K\downarrow$ states and the contribution from the $K\uparrow$ state is vanishing. The measured current originates thus from a single valley with single spin component. For the peak of the total ANC, the contribution from the $K$ valley is dominating. Therefore, with lowering the Fermi level, single spin current (down spin) carrying single valley information can be generated by a temperature gradient, which hints at possible applications in spin caloritronics and valleytronics.

This work is supported by Hundred Talents Program of The Chinese Academy of Sciences. G. S. is supported in part by the MOST (Grant Nos. 2012CB932900, 2013CB933401), the NSFC (Grant No. 11474279) and the CAS (Grant No.XDB07010100).  The Center for Nanostructured Graphene (CNG) is sponsored by the Danish National Research Foundation, Project DNRF58.
%\end{acknowledgments}

%\begin{Appendix}
%\newpage
%\section{Supplementary materials}

%\end{Appendix}

\end{document}